# Multi-Level Association Rule Mining for Wireless Network Time Series Data

Chen Zhu[1*], Chengbo Qiu[1], Shaoyu Dou[1], Minghao Liao[1]

[1]Department of Computer Science and Technology, Tongji University, Shanghai, 201804, China

[*]Corresponding author e-mail: 2030802@tongji.edu.cn

**Abstract.** Key performance indicators(KPIs) are of great significance in the monitoring of wireless network service quality. The network service quality can be improved by adjusting relevant configuration parameters(CPs) of the base station. However, there are numerous CPs and different cells may affect each other, which bring great challenges to the association analysis of wireless network data. In this paper, we propose an adjustable multi-level association rule mining framework, which can quantitatively mine association rules at each level with environmental information, including engineering parameters and performance management(PMs), and it has interpretability at each level. Specifically, We first cluster similar cells, then quantify KPIs and CPs, and integrate expert knowledge into the association rule mining model, which improve the robustness of the model. The experimental results in real world dataset prove the effectiveness of our method.

## 1. Introduction

AIOps aims to use various operation and maintenance data, such as log data and KPI time series data, to help engineers further analyse and make decisions on problems through machine learning methods. It has good research and application prospects[1]-[3]. In the field of operation and maintenance of wireless network data transmission system, the distribution of users in the cells covered by the base stations is complex and dynamic, which brings great challenge against wireless system maintenance and optimization[4].

In this paper, we propose a multi-level association rule mining method to effectively mine associations between the CPs and KPIs. he obtained rules can be employed for network resource optimization [5]-[8]. In addition, the proposed method has significant advantages over traditional association analysis methods. Our method can qualitatively and quantitatively analyse the association between CPs and KPIs. The precision of more than 90% is achieved in the real dataset, which will effectively help engineers adjust the CPs of base stations to make wireless network service quality reach the desired level.

## 2. Related work

The existing association analysis approaches can be divided into the following three categories: similarity-based approaches, correlation-based approaches and graph model-based approaches.

### 2.1. Similarity-based approaches

By directly calculating the distance or similarity between variables, common algorithms include classic k-means[9], dynamic time warping (DTW) model, and alarm2vec algorithm. DTW model is a common method to measure the similarity of time series data. Based on the idea of dynamic programming, it is used to determine the similarity between two series by solving the regularization function that satisfies the minimum cumulative distance between two series. The Alarm2vec model is based on the node2vec model, which maps different types of alarms into the hidden space for similarity calculation.

*2.2. Correlation-based approaches*

According to different data types, it can be divided into correlation analysis between time series[10], correlation analysis between event series[11], and correlation analysis between event series and time series[12]. The correlation between time series is estimated using correlation analysis methods, such as Pearson coefficient, Spearman coefficient and Kendall coefficient. The correlation algorithms between event sequences can be divided into alarm correlation algorithms based on similarity measurement, alarm correlation algorithms based on knowledge base, and event correlation algorithms based on statistical information. Frequent item mining algorithms are commonly used, such as Apriori[13] and FP-growth[14]. Apriori was first proposed by Agrawal. It uses multiple iterations to establish candidate sets to find frequent items. The frequent pattern growth algorithm is proposed for frequent item mining using frequent pattern trees. It is a computational efficiency and scalability algorithm for frequent item mining, and is superior to other association rule algorithms in many cases. Compared with Apriori, FP-growth only traverses data twice, and does not need to generate candidate sequences, which greatly improves the mining efficiency.

*2.3. Graph model-based approaches*

The association analysis approaches based on graph model aim to characterize the association between variables by constructing a graph model between variables and using the edges between nodes. Literature[15] proposed a deep learning model TCDF based on attention mechanism, through which a sequence with strong association with the target sequence can be obtained. In addition to the above method of constructing the graph model of the relationship between variables based on the regression model between learning sequence pairs, there is also a common method of association algorithm based on causal analysis, that is, learning the causal relationship between variables and using directed acyclic graph model (DAG). Literature[16] proposed a constraint based PCMCI method, which is a two-stage causal analysis method. The first stage is the discovery process of association, that is, the conditional independence test is used to iteratively learn the possible parent node (dependent variable) for the target variable and the second stage is the orientation process of the causal direction, that is, the edges obtained in the first stage are judged by using such rules as time sequence and d-separate, so that a DAG can be obtained. Nodes in the graph represent different variables, and directed edges are represented by the direction from dependent variables to outcome variables. Literature[17] proposed the variant LPCMCI of PCMCI to discover the causal relationship of the same period.

**3. Proposed method**

*3.1. Problem statement*

The goal of our method is to build an association model, based on the environmental information $s$, the CPs $a$, and the KPIs $y$, according to the historical data. Given $s$, output the specific changes of the associated KPIs, including the direction and amplitude, according to the adjusted $a$. For example, given $s$, if the adjusted $a$ is the electronic dip angle, when the electronic dip angle is increased or decreased, the change of associated $y$ such as the number of users, downlink traffic, average throughput needs to be output, including the direction and amplitude. According to the understanding of the problem, we propose the following theoretical model:

$$\mathbf{y}(t) = f(\mathbf{a}(t), \mathbf{s}(t), \mathbf{d}(t), \mathbf{y}(t-1)), \tag{1}$$

$\mathbf{y}(t)$ are the values of the KPIs at time t, $\mathbf{a}(t)$ are the values of the CPs adjusted at time t, $\mathbf{s}(t)$ is the environmental information, $\mathbf{d}(t)$ is the adaptive mechanism and can be regarded as an unobservable random process, which may have a certain impact on the KPIs.

Furthermore, the above equation can be transformed into a problem of association analysis with noise and unobservable variables:

$$y_j(t) = f_j(pa(y_j(t)), \eta_j(t)), \tag{2}$$

$$pa(y_j(t)) \in \{\mathbf{a}(t), \mathbf{s}(t), \mathbf{d}(t), y_j(t-1)\}, \tag{3}$$

$y_j(t)$ is the value of the number $j$ KPI at time t and $\eta_j(t)$ is the noise.

Based on the above problems, there are mainly the following research difficulties:
1) Hidden variables: there are hidden variables in the sample data, such as some environmental variables, the adaptive mechanism of the base station, etc. These variables are difficult to measure in practice and will affect the analysis of some association rules.
2) Noise: sample data may generate noise during collection and aggregation.
3) Large sampling space: the high dimensionality of CPs and KPIs makes the data sampling space large.
4) Multi-level: the same variable may have different association rules under different levels.
5) Interpretability: the method needs to have interpretability for engineers to verify.

### 3.2. General framework
Our framework is shown in Figure 1. Given the environmental information, the qualitative relationship between CPs and KPIs of the target is mined, and the impact of the adjustment of CPs on KPIs is quantitatively expressed. The first stage is hierarchical clustering of similar cells, as shown in Figure 2. The second stage is our multi-level association rule mining model. The third stage is based on feature selection, adding some important environmental features to the association rules in the second stage, and the last stage is the output of the association rules.

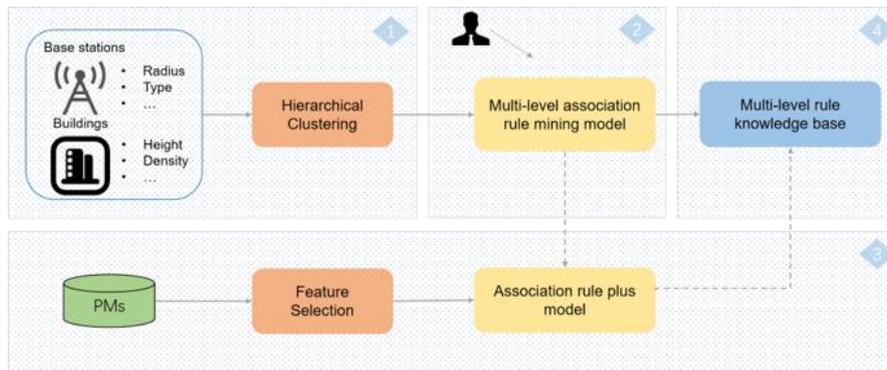

**Figure 1.** The overall framework of our method.

### 3.3. Preprocessing module
The preprocessing of the original wireless network data (KPIs, CPs, PMs and other data) mainly including redundant data filtering, missing data completion and data quantification. Considering the robustness requirements of the method, the performance of the method needs to be guaranteed in the case of inaccurate measurement of some data and noise in the data. According to the quality of experience(QoE), the KPIs have little impact on the user experience within a certain range. As shown in Figure 3, the KPIs are divided into several quantitative levels.

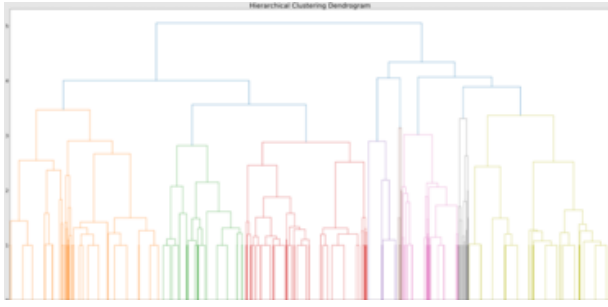 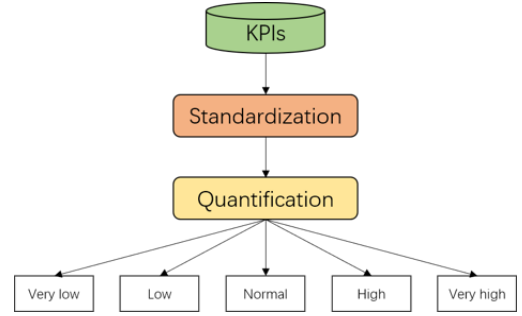

**Figure 2.** Hierarchical clustering of similar cells.     **Figure 3.** KPIs quantification based on QoE.

*3.4. Multi-level association rule mining model*

In our framework, after the data is constructed into a transaction database, we can use multi-level association rule mining method to effectively mine the association rules of different levels between CPs and KPIs, and support, confidence and lift are provided to filter and sort the rules:

$$Support_{B \to A} = P(AB), \tag{4}$$

$$Confidence_{B \to A} = \frac{P(AB)}{P(B)}, \tag{5}$$

$$Lift_{B \to A} = \frac{P(AB)}{P(A)P(B)}. \tag{6}$$

Set the KPI as $A = \{a^1, ..., a^l, ...a^L\}$, and the CP as $B = \{b^1, ..., b^k, ...b^K\}$, $N_{all}$ is the total number of transaction items, $N_A, N_B$ are the number of occurrences of KPI A and CP B respectively, $N_{AB}$ is the number of occurrences of A and B. The quantitative relationships between different levels are described as follows:

$$Support_{B \to A} = \frac{N_{AB}}{N_{all}} = \frac{\sum_{l,k} N_{a^l b^k}}{N_{all}} = \sum_{l,k} Support_{b^k \to a^l}, \tag{7}$$

$$Confidence_{B \to A} = \frac{N_{AB}}{N_B} = \frac{\sum_{l,k} N_{a^l b^k}}{N_B} = \sum_{l,k} \left(\frac{N_{a^l b^k}}{N_{b^k}} \cdot \frac{N_{b^k}}{N_B}\right)$$
$$= \sum_k \frac{N_{b^k}}{N_B} \sum_l \frac{N_{a^l b^k}}{N_{b^k}} = \sum_k \frac{N_{b^k}}{N_B} \sum_l Confidence_{b^k \to a^l}, \tag{8}$$

$$Lift_{B \to A} = \frac{N_{AB} N_{all}}{N_A N_B} = \frac{N_{all} \sum_{l,k} N_{a^l b^k}}{N_A N_B} = \sum_{l,k} \left(\frac{N_{a^l} N_{b^k}}{N_A N_B} \cdot \frac{N_{all} N_{a^l b^k}}{N_{a^l} N_{b^k}}\right)$$
$$= \sum_{l,k} \left(\frac{N_{a^l} N_{b^k}}{N_A N_B} \cdot \frac{P(a^l b^k)}{P(a^l) P(b^k)}\right) = \sum_{l,k} \frac{N_{a^l} N_{b^k}}{N_A N_B} Lift_{b^k \to a^l}. \tag{9}$$

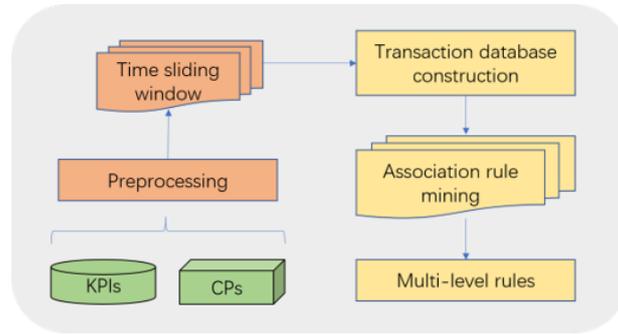

**Figure 4.** Framework of multi-level association rule mining model.

*3.5. Association rule plus model*
The association rule plus model takes some important environmental features such as user density as input through feature selection of PMs, and constructs candidate supersets for each rule in the output result of the second stage combined with the input $s$, and prunes according to the closure principle: if there is an association rule whose support is greater than or equal to the threshold value, that is, $Sup(I) \geq t$, then there must be $Sup(i) \geq t$ for any non-empty subset $i$ of $I$. Finally, the support of candidate supersets is calculated, and the rule sets containing CPs $a$, environment information $s$ and KPIs $y$ are obtained after filtering and splicing.

## 4. Experiments

*4.1. Datasets*
Our data is based on the real time series data monitored and collected by the 4G base stations and the cells they serve. A total of 2646 base stations and cells data were recorded in 2020.11.10-2020.11.21. The data mainly includes the engineering parameters, CPs, PMs, and KPIs of the base stations. The engineering parameters refer to the relevant parameters during the planning and construction of the base station. PMs characterize network quality, including energy consumption, load, traffic and other indicators. The CPs refer to the description information of the base stations. There are 22 key CPs and 25 KPIs that we focus on.

*4.2. Baselines*
We use the PCMCI algorithm as the baseline. The PCMCI algorithm is a constraint-based time series causal analysis model, which judges the causal relationship of variables by analysing the conditional independence between variables. According to the labels of some association rules obtained from expert knowledge, we use the precision as the evaluation index to compare PCMCI with our method.

At the same time, in order to test the robustness of our method to noise, we added noise to the original data for association rule mining and compared it with the experimental results without noise. If there are fewer redundant rules after adding noise, it indicates that our method has good robustness.

Finally, since the heights of the buildings around the base stations have an impact on the association rules, in order to evaluate the applicability of our method to different scenarios, we conducted comparative experiments on different building heights.

*4.3. Results*
Table 1 shows the precision comparison between our method and PCMCI. It can be seen that our method is superior to the baseline method in three of the four cells and our method has better average performance.

**Table 1.** Comparison of precision between PCMCI and our method.

| Cell | PCMCI | Our method |
|---|---|---|
| Cell 1 | 0.8900 | **0.9200** |
| Cell 2 | 0.7000 | **0.8947** |
| Cell 3 | 0.6000 | **0.9474** |
| Cell 4 | **1.0000**(4 rules only) | 0.8800(25 rules) |
| average | 0.8076 | **0.9122** |

To verify the robustness of the method, we randomly select 20% of the KPI data of all cells to add interference. The interference noise is sampled from the uniform distribution, and the noise amplitude is 10% of the original KPI standard variance. The association rules are mined on five different cell clusters. Table 2 shows the minimum number of top k support rules in the original data that all appear in the rules with added noise. The closer the value is to k, the better the robustness of the method.

**Table 2.** Robustness test of our method after adding noise.

| Cluster | top 10 rules | top 15 rules | top 20 rules |
|---|---|---|---|
| Cluster 1 | 13 | 21 | 25 |
| Cluster 2 | 18 | 20 | 25 |
| Cluster 3 | 14 | 25 | 32 |
| Cluster 4 | 13 | 19 | 26 |
| Cluster 5 | 12 | 22 | 29 |
| average | 14.0 | 21.4 | 27.4 |

The height of surrounding buildings will affect the base station, resulting in different association rules. Table 3 shows the results of our method for mining association rules in different scenarios. When other parameters remain unchanged, the CPs in the table can be reduced to improve the coverage performance of the system broadcast. We can see that our method can flexibly obtain association rules in different scenarios.

**Table 3.** Association rule mining in different scenarios.

| Building height(meters) | Station height(meters) | CP | KPI |
|---|---|---|---|
| 11.30 | 18 | CELLSIMAP.SITRANSECR $\rightarrow$ 2 | rrc setup success rate $\rightarrow$ normal |
|  |  | CELLSIMAP.SITRANSECR $\rightarrow$ 4 | rrc setup success rate $\rightarrow$ very low |
| 35.78 | 23 | CELLSIMAP.SITRANSECR $\rightarrow$ 2 | rrc setup success rate $\rightarrow$ low |
| 43.65 | 35 | CELLSIMAP.SITRANSECR $\rightarrow$ 2 | rrc setup success rate $\rightarrow$ very low |

## 5. Conclusion
In this paper, we propose a framework for mining association rules in wireless networks. We use hierarchical clustering to combine highly similar cells into cell clusters, use a multi-level association rule mining model to obtain the relationship between CPs and KPIs, and give parameter adjustment suggestions when the network quality doesn't reach expectation. When the model is applied to target tasks, it can not only analyse qualitatively, but also mine quantitative association rules. At the same time, the complexity of multi-level mining is low, and it is interpretable. The experimental results on real world dataset show that the performance of our method is better than baseline method, it is also robust to noise and suitable for different scenarios. In the future, we may consider to extend the proposed approaches to a distributed setting which allows for parallel processing and can protect users' privacy better[18]-[19].